\documentclass[11pt]{article}
\usepackage{moriond2000,epsfig}

\def\be{\begin{equation}}
\def\ee{\end{equation}}
\def\bea{\begin{eqnarray}}
\def\eea{\end{eqnarray}}

\begin{document}
\vspace*{4cm}
\title{STUDY OF THE DEPLETION EFFECT IN THE CLUSTER MS1008-1224}

\author{C. MAYEN, G. SOUCAIL}

\address{Observatoire Midi-Pyr\'en\'ees, Laboratoire d'Astrophysique,
UMR 5572, 14 Avenue E. Belin, 31400 Toulouse, France}

\maketitle\abstracts{ We present a detailed study of the depletion
effect (the radial manifestation of the magnification bias) in the 
cluster MS1008-1224.
Following our results concerning the simulations of depletion curves (see
Mayen and Soucail, 2000 \cite{moi} for a complete description of the code), we propose to 
constrain the mass profile of the cluster and the main characteristics
of its potential (orientation and ellipticity). This application is
based solely on deep photometry of the field and does not require the
measurement of the shape parameters of the faint background galaxies.} 


\section{Introduction}

When the logarithmic slope of the galaxy counts is lower than 0.4 (this
is the case in all filters at large magnitude), the
magnification bias due to a gravitational lens makes 
the number density of objects
decrease, and consequently,
the radial distribution shows a typical depletion curve.
This effect results from the competition between the gravitational
magnification that increases the detection of individual objects and
the deviation of light beam that spatially magnifies the observed area
and thus decreases the apparent number density of sources. Since a few years, several tentative 
analysis have been proposed 
to determine cluster mass distribution (Broadhurst et al., 1995 \cite{broadhurst}; Taylor 
et al., 1998 \cite{taylor}), the redshift distribution of sources and to bring constraints on
cosmological parameters (Fort et al., 1997 \cite{fort}). 
 
Here, we propose to use depletion curves in order to
constrain the mass profile of MS1008-1224 and some characteristics 
of its potential (orientation and ellipticity) by combining 
results of simulations of depletion curves obtained with 
different lens models (Mayen and Soucail, 2000 \cite{moi}) 
and high quality multicolor images of MS1008-1224 obtained with FORS and ISAAC 
during the science verification phase of the VLT-ANTU (UT1) at Cerro
Paranal. It is a
very rich galaxy cluster, located at $z=0.3062$ (Lewis et al., 1999 \cite{lewis}), slightly 
extended in X-rays, with a galaxy
distribution quite circular, surrounding a North-South elongated core.
There is a second clump of galaxies to the North. Some gravitationally
lensed arcs to the North and the East of the field have been reported by Le F\`evre et al., 1994 
\cite{lefevre} as well as by Athreya et al., 2000 \cite{athreya} who also
detected a high redshift cluster lensed by MS1008-1224 in the South-West
part of the field using photometric redshifts. 

In practice, we used only the FORS data which extend to a larger
distance and are more suited for our study. Background 
sources were selected by removing the elliptical galaxies  
sequence in a color-magnitude diagram ($R-I\ vs \ R$), essentially 
valid at relatively bright
magnitudes. This statistical correction is not fully reliable as it does
not eliminate bluer cluster members or foreground sources.

Our counts may also present
an overdensity of objects in the inner part of the cluster
($r<20''$) due to the non-correction of the surface of cluster
elliptical galaxies we have removed. This effect is more sensitive in
the inner part of the cluster where these galaxies are dominant. But
as it is not in the most interesting region of the depletion area, we
did not try to improve the measures there.

In addition, due to the presence of two 11 magnitude stars in the
northern part of the FORS field, two occulting masks were put to avoid
excessive bleeding and scattered light. We took into account this
partial occultation of the observed surface for the radial counts
above a distance of 130''. This may
possibly induce some additional errors in the last two points of the
curves, which are probably underestimated, because of the difficulty
to estimate the surface of the masks and some edge effects at the
limit of the field.


\section{Depletion curves and mass density profile}

The first step would be to locate the cluster center. Its position
is rather difficult to estimate directly from the
distribution of the number density of background sources, although in
principle one should be able to identify it as the barycenter of the
points with the lower density around the cluster. We did not
attempt to fit it and preferred to fix it 15''
North of the cD, following both the X-ray center position (Lewis et al., 1999
\cite{lewis}) or the weak lensing center (Athreya et al., 2000) \cite{athreya}.

\begin{figure}
\centerline{
\psfig{figure=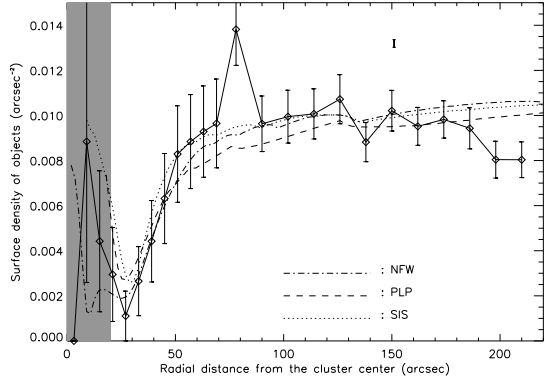,height=5.4cm}}
\caption{Depletion curve measured in MS1008-1224 in the I band. The
best fit by the three models described in text are
given. Error bars correspond to Poisson statistical noise. In the shaded area, the
signal cannot be constrained observationally (decrease in the observed area, obscuration 
by the brightest cluster galaxies ...).}
\label{depms1008}
\end{figure}

The radial counts were performed in a range of 3 magnitudes up to the
completeness magnitude in the B, V, R and I bands and up to a radial
distance of 210'' from the center, which covers the entire FORS
field. The depletion was clearly detected in the four bands (see 
Mayen and Soucail, 2000 \cite{moi} and Figure \ref{depms1008}).    
The count step was fixed to 30 pixels (6'') in the
innermost 80'' and for the rest of the field we adopted a count
step of 60 pixels (12'') in order to reduce the statistical error
bars. These values are a good balance between statistical errors in
each bin which increase for small steps and a reasonable spatial
resolution in the radial curve, limited by the bin size.

We fitted the observed depletion curve in the I band with three 
mass models (Figure \ref{depms1008}), namely a singular isothermal
sphere (SIS), a power-law density and a Navarro, Frenk \& White \cite{navarro} 
(NFW) profiles. For each model, a 
$\chi^2$ minimization was introduced to derive the best fit and
their related parameters (Table \ref{chi2table}). The absolute normalization of 
the mass profiles results from the count model in empty field we have used (which 
fix the redshift distribution of the background sources before magnification by the 
gravitational lens). The curve was fitted after removing some clear deviant points : the  
last two points probably poorly corrected from edge effects, and
those associated to the overdensity seen at $r \simeq 80''$. 
This bump is easily identifiable in the V, R and I
curves, and can be partly explained by the presence of a background
cluster lensed by MS1008-1224 and identified by Athreya et al., 2000 
\cite{athreya}. Nevertheless, even if we remove from our data all the
lower right quadrant of the field where this structure is located, the
bump is still there although significantly reduced. This suggests that
it may be more extended behind the cluster center than initially
suspected.

\begin{itemize}
\item[$\bullet$] The velocity dispersion derived from the SIS model
($\sigma_{\textrm{fit}} = 1200 \pm ^{200}_{175}$ km s$^{-1}$) is in good
agreement with the value measured by Carlberg et
al., 1996 \cite{carlberg} ($\sigma_{\textrm{obs}} = 1054 \pm 107$ km s$^{-1}$) but is
more discordant with the value of 900 km s$^{-1}$ inferred from the
shear analysis of Athreya et al., 2000 \cite{athreya}.
\item[$\bullet$] The slope of the potential fit with a power-law
density profile is close to an isothermal one ($\alpha = 1.88 \pm ^{0.27}_{0.25}$),
although slightly shallower.
\item[$\bullet$] For a NFW profile, we find a virial radius
($r_{200}=3.2 h_{50}^{-1}$ Mpc) and a concentration parameter
($c=8.9$) quite in good agreement with those of Athreya et al., 2000 
\cite{athreya} derived from weak lensing measures.
\item[$\bullet$] The comparison between the 3 fits favors a NFW
profile as the best fit of our depletion curve (Table
\ref{chi2table}), also in agreement with the shear results for this
cluster.
\end{itemize}

\begin{figure}
\centerline{
\psfig{figure=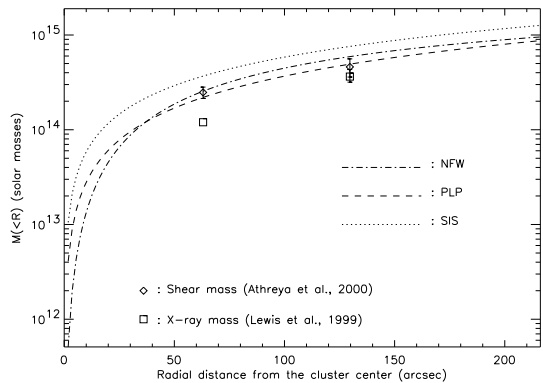,height=5.4cm}}
\caption{Mass profiles of MS1008-1224 inferred from the depletion 
analysis. X-ray and shear masses are also shown. The agreement 
between the two lensing mass estimates is very good.}
\label{profilmasse}
\end{figure}

It is important to note that the mass profiles inferred from the depletion 
analysis are in good agreement with the shear masses (Figure
\ref{profilmasse}). Thus, the masses found by using two independent lensing mass
estimates are consistent whatever the radial distance from the cluster
center. The discrepancy with X-ray mass still remains, although it is
slightly reduced at large distance from the center. This encouraging result allows to 
consider the use of the depletion as a secure alternative solution against the 
other lensing mass estimates when the quality of the data is not sufficient to 
allow the use of the usual ones. 

The study of the shape of the depletion area, which can be easily related to the 
ellipticity and the orientation of the mass distribution, has also allowed us to constrain 
these parameters with a good accuracy for MS1008-1224 (see Mayen and Soucail, 2000 
\cite{moi}). 

\begin{table}
\caption{Results of the fit of the model parameters. 
The errors are given at the 99.9 \%\ confidence level.}
\label{chi2table}
\vspace{0.4cm}
\begin{center}
\begin{tabular}{ccc}
\hline\noalign{\smallskip}
Model & Parameter & reduced $\chi^2$ \\
\hline\noalign{\smallskip}
SIS & $\sigma=1200\pm^{200}_{175}$ km s$^{-1}$ & 0.49 \\
PLP & $\rho_E=2.4\pm^{1.3}_{0.6} \times 10^{15}$ M$_{\odot}/$Mpc$^3$
& 0.62 \\
PLP & $\alpha=1.88\pm^{0.27}_{0.25}$ & 0.65 \\
NFW & $r_{200}=3.2\pm^{0.7}_{0.6}$ Mpc & 0.44 \\
NFW & $c=8.9\pm^{6.7}_{4.0}$ & 0.45 \\
\noalign{\smallskip}\hline
\end{tabular}
\end{center}
\end{table}

\section{Conclusion and outlook} 

The study of the depletion effect in the cluster MS1008-1224 with very deep and 
high quality VLT images has allowed us to constrain the mass profile up to a 
reasonable distance from the center and to constrain the ellipticity and 
the orientation of the mass distribution with a good accuracy. The results found 
are consistent with those inferred from other lensing techniques. Thus, the 
depletion reveals itself as a secure technique. The next step to explore is the 
influence of the clustering of background sources and the interest of wide field 
imaging : The normalization of the field  
number counts for the magnification bias can also be estimated outside
the cluster, in exactly the same observing conditions (filter,
magnitude limit, seeing, \ldots ), giving an absolute calibration of
the depletion effect. A more prospective point which still requires to be studied in details 
is the reconstruction of two-dimensional mass maps of clusters from the 
depletion signal only.

\section*{Acknowledgments}

We wish to thank the European TMR Programme \textit{Gravitational
Lensing : New Constraints on Cosmology and The Distribution of Dark
Matter} for its financial support.



\end{document}